\documentstyle[preprint,aps,epsfig]{revtex}
\tighten
 \newcommand \be {\begin{equation}}
\newcommand \bea {\begin{eqnarray} \nonumber }
\newcommand \ee {\end{equation}}
\newcommand \eea {\end{eqnarray}}
 
 \newcommand \bi {\bibitem}

\newcommand \lan {\langle}
\newcommand \ran {\rangle}

\newcommand \Tr {\mbox{Tr}}
\newcommand{\go}{\rightarrow}

\begin{document}
\draft
\preprint{Preprint no.}
\title{On the evaluation of the specific heat and
       general off-diagonal n-point correlation functions
       within the loop algorithm}

\author{J.V. Alvarez and Claudius Gros}

\address{Fachbereich Physik, Universit\"at des Saarlandes,
         Postfach 151150, 66041 Saarbr\"ucken, Germany}

\date{\today}
\maketitle

\begin{abstract}
 
We present an efficient way to compute diagonal and
off-diagonal n-point correlation functions for 
quantum spin-systems 
within the loop algorithm. We show that the general rules
for the evaluation of these correlation functions
take an especially simple form within the framework of
directed loops. These rules state that contributing
loops have to close coherently. As an application
we evaluate the specific heat for the case
of  spin chains and  ladders.
\end{abstract} 

\vfill
\pacs{64.70.Nr, 64.60.Cn}
%{\bf \hfill cond-mat/9502045}

\vfill
%\newpage

%\baselineskip 6mm
\narrowtext

%%%%%%%%%%%%%%%%%%%%%%%%%%%%%%%%%%%%%%%%%%%%%%%%%%%%%%%%%%
%%%%%%%%%%%%%%%%%%%%%%%%%%%%%%%%%%%%%%%%%%%%%%%%%%%%%%%%%%

\section{Introduction}

Numerical investigations of strongly correlated electron
systems \cite{Dagotto} gained considerable importance
in the last decade. The evaluation of non-diagonal
correlation function and dynamical response function
plays a major role in the context of correlated
electron systems \cite{Dagotto,dyn_response}. On the
other hand, there are only very few investigations of
non-diagonal and/or higher-order
correlation function in the context of
quantum spin-systems. Indeed, it has been realized only
recently, that non-diagonal correlation function might
be calculated efficiently within the loop-algorithm
\cite{WIESE}. The loop-algorithm \cite{EVERTZ} has 
established itself as the method of choice for 
quantum-Monte Carlo (MC) simulations of non-frustrated
quantum spin systems. 

The key observation here is the fact, that local
updating dynamics in a MC simulation creates
strongly correlated configurations for 
gapless quantum spin systems at low temperatures. 
Since the samples are then not statistically
independent, the statistical error bars do decay
only very slowly with the number of samples.
One way to state this problem is to say, that
the autocorrelation time $\tau_{auto}$ for the samples
of spin-configurations created with the MC-walk
increases (in generally exponentially) at
low temperatures.

Most efficient MC procedures implement consequently 
global update dynamics. Examples of these procedures are the 
clusters algorithms \cite{SWENDSEN}.
Designed to circumvent the critical slowing down, these   
methods  have been intensively used to study classical statistical 
systems near critical points, where the problem of large
$\tau_{auto}$ is very severe.  
    
The loop algorithm \cite{EVERTZ} can be considered 
as a generalization  of classical clusters algorithms 
to quantum models. In fact, it gives a 
prescription on how global updates can  be performed 
in quantum systems.
As we will see this prescription lays on  the geometric interpretation 
of the transformation from a  quantum system to a statistical 
model of oriented loops. The MC procedure can be implemented then
directly on the loops. 
It has the advantage that the updating dynamics 
defined on the loops generates statistically nearly independent
configurations. The autocorrelation time is therefore about
just MC step  and the corresponding  operators can be measured 
at every MC step avoiding both 'waiting times' 
and substantial increments of the variance (statistical
error bars).    

In addition, a loop has another remarkable property;
starting from an allowed spin configuration, 
constructing a loop and then flipping all spins in one loop 
(flipping the orientation of the loop) 
one obtains a new allowed configuration.  
This observation allows to compute the expectation value
of operators not only in one configuration per MC step 
but in all configuration related to it 
by  flipping any number of given loops.
This procedure is usually called improved estimator \cite{WIESE}.

The purpose of this work is to extend the algorithm 
to the computation of higher order 
(and non-diagonal) correlations functions.
As we will see it involves dealing with two or more loop 
contributions. In particular we will focus on the 
specific heat  $c_V$, which, in the past,
has been considered
a major challenge for Monte-Carlo simulations
\cite{Huscroft}. We will show, that the direct 
evaluation of the higher-order (non-diagonal)
correlation functions contributing to $c_V$
allows for improved estimators and such to gain
one order of magnitude in computational efficiency.
The method that we presented is valid  in 
any dimension.

%%%%%%%%%%%%%%%%%%%%%%%%%%%%%%%%%%%%%%%%%%%%%%%%%%%%%%%%%%
%%%%%%%%%%%%%%%%%%%%%%%%%%%%%%%%%%%%%%%%%%%%%%%%%%%%%%%%%%

\section{The  Loop Algorithm}  
  
A nice review of the loop algorithm can be found in
Ref.\ \cite{EVERTZ_rev}. Here we start with a short introduction
in order to introduce the notation used further on for the
evaluation of higher-order correlation functions.

The loop algorithm is most easily understood in the checkerboard
picture for a discrete number of Trotter slices $N_T$; the generalization
to continuous Trotter time \cite{cont_time} 
is straightforward. 
This picture, which is based on the Suzuki-Trotter
decomposition, describes in a graphical way how  
the interacting spin system  wave function evolves in 
discrete imaginary time.
   
The Suzuki-Trotter formula \cite{Trotter}
maps a quantum spin system in dimension $d$
onto a classical spin in dimension $d+1$. The partition function 
of the original quantum spin model is hereby written
in terms of the trace of a product of 
transfer matrices defined in the classical model.
              
To illustrate the method we consider an inhomogeneous 
one-dimensional XXZ model $H=H_1+H_2$ on a bipartite chain
of length $L$: 

\[
H_{1}=\sum_{i=2m}  H_{i},\qquad 
H_{2}=\sum_{j=2m+1}  H_{j}  
\]
\[ 
H_{i}=-\frac{J_{i}^{XY}}{2}
\left(S_{i}^{+}S_{i+1}^{-}+S_{i}^{-}S_{i+1}^{+}\right)
+J_{i}^{Z}S_{i}^{Z}S_{i+1}^{Z}~,
\]
where the sign of the term $\sim J_{i}^{XY}$ has been choose
to be negative by an appropriate rotation of the spins
on one of the two sublattices. This is always possible on a
bipartite lattice and allows for positive transfer matrix elements 
(absence of the sign problem). The decomposition
$H=H_1+H_2$ allows for the use of Totter-Suzuki formula
\cite{Trotter} for the representation of the partition
function $Z= \Tr \left[\exp(-\frac{\beta}{N_{T}}H)\right]^{N_{T}}$,

\[
Z=\Tr \prod_{n=1}^{N_{T}}\sum_{\alpha_n}
 \lan \phi_{\alpha_n}^{(n)}|
\exp(-\Delta\tau H_1) \exp(-\Delta\tau H_2)
| \phi_{\alpha_{n+1}}^{(n+1)}\ran + O({\Delta\tau^2})~, 
\]
where $\Delta\tau=\beta/N_T$. Here we have introduced
representations of the unity operator
$\sum_{\alpha_n}|\phi_{\alpha_n}^{(n)}\ran
\lan \phi_{\alpha_n}^{(n)}|$ in between any of the
$N_T$ imaginary time slices.

Since $H_{1}$ and $H_{2}$ are sum of local operators
that commute with each other, we may write 
the wave function as the  product of the local basis
in say z-component of spin,
$|\phi_{\alpha_n}^{(n)}\ran=\otimes_i
|\sigma_i\ran$, with
$\sigma_i=\uparrow,\downarrow$.

\begin{figure}[thb]
\begin{center}

\epsfig{file=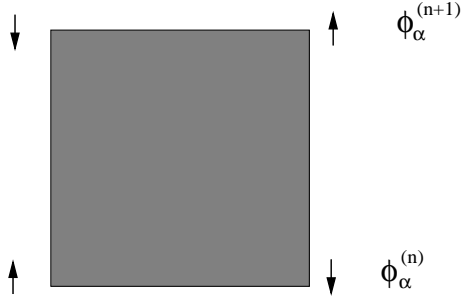,  height=6cm,
width = 4cm, angle=-90}
\medskip

\caption[]{Illustration of a plaquette with a spin flip process 
           which corresponds to the matrix element 
$ \lan \uparrow \downarrow | \exp(-\Delta\tau H)
|\downarrow \uparrow \ran $ of the transfer matrix.}
\label{plaquette}
\end{center}
\end{figure}

In the checkerboard lattice the interaction between two consecutive
pairs of spins
is graphically denoted by shaded plaquettes 
(see Fig.\ \ref{plaquette}). 
There are two spins interacting per plaquette so a 4x4 transfer matrix
$T_i$ can be defined in each plaquette, which depends only
one the coupling constants.

%%%%%%%%%%%%%%%%%%%%%%%%%%%%%%%%%%%%%%%%%%%%%%%%%%%%%%%%%%%%%%%%%%%%%%%
%%%%%%%%%%%%%%%%%%%%%%%%%%%%%%%%%%%%%%%%%%%%%%%%%%%%%%%%%%%%%%%%%%%%
   
For the XXZ-model the transfer matrix $T_i$ is in the
basis $\left( |\uparrow,\uparrow\ran,
|\uparrow,\downarrow\ran,|\downarrow,\uparrow\ran,
|\downarrow,\downarrow\ran\right)$:
\[
T_i= \mbox{e}^{{\Delta\tau J_i^z\over4}}
\left( \begin{array}{cccc}  
\exp(-\frac{\Delta\tau J^{Z}_{i} }{2})  &   0    &  0     & 0 \\
0  & \cosh(\frac{\Delta\tau J^{XY}_{i}}{2})
&\sinh(\frac{\Delta\tau J^{XY}_{i}}{2}) & 0 \\
 0  & \sinh(\frac{\Delta\tau J^{XY}_{i}}{2})& 
\cosh(\frac{\Delta\tau J^{XY}_{i}}{2}) & 0 \\
  0  &    0   &   0    &  \exp(-\frac{\Delta\tau J^{Z}_{i}}{2}) 
                \end{array} \right)~. 
\]
The partition function $Z$ is then, up to terms order 
$O(\Delta\tau^2)$,
the trace of a product of transfer matrices:  

\[
Z=\Tr\left[\exp(-\beta H)\right]= 
\Tr \prod_{n=1}^{N_{T}}
\left( \bigotimes_{i=2m} T_{i}\right)
\left( \bigotimes_{j=2m+1} T_{j }\right)~. 
\] 
As a next step beyond this standard representation of
$d$-dimensional quantum models
in terms of classical statistical systems \cite{Beard_96}
we expanded the transfer matrices 
$T_i=\sum_{\gamma} p_{i}^{(\gamma)}M^{(\gamma)}$
in terms of certain matrices $M^{(\gamma)}$ such that the
weight $p^{(\gamma)}_{i} \ge 0$ are non-negative. This
is, in general, not possible for all models.
For the XXZ with $J^{XY}_{i}\ge J_{i}^{Z}$ we can choose:

\[   
M^{(1)}=\left( \begin{array}{cccc}            
              1  &   0    &  0   & 0 \\   
              0  &   1    &  0   & 0 \\
              0  &   0    &  1   & 0 \\
              0  &   0    &  0   & 1 
                \end{array} \right), \quad 
       M^{(2)}=\left( \begin{array}{cccc}            
              1  &   0    &  0   & 0 \\   
              0  &   0    &  1   & 0 \\
              0  &   1    &  0   & 0 \\
              0  &   0    &  0   & 1 
                \end{array} \right), \quad
       M^{(3)}=\left( \begin{array}{cccc}            
              0  &   0    &  0   & 0 \\   
              0  &   1    &  1   & 0 \\
              0  &   1    &  1   & 0 \\
              0  &   0    &  0   & 0 
                \end{array} \right)~,
\]
where
$p_{i}^{(1)}=\frac{1}{2}(\exp(-\Delta\tau J^z_i/2)
                        +\exp(-\Delta\tau J^{XY}_{i}/2)) 
\exp(\Delta\tau J^{Z}_{i}/4)$,
$ p_{i}^{(2)}=\frac{1}{2}(\exp(-\Delta\tau J^z_i/2)
                         -\exp(-\Delta\tau J^{XY}_{i}/2)) 
\exp(\Delta\tau J^{Z}_{i}/4)$ and
$p_{i}^{(3)}=\frac{1}{2}(-\exp(-\Delta\tau J^z_i/2)
                        +\exp(\Delta\tau J^{XY}_{i}/2)) 
\exp(\Delta\tau J^{Z}_{i}/4)$.  
We then obtain for the partition function

\be
Z=\Tr \prod_{n=1}^{N_{T}} \bigotimes_{i=2m}\left(
\sum _{\gamma} p_{i}^{(\gamma)}M^{(\gamma)} \right)
\bigotimes_{j=2m+1}\left(
\sum _{\gamma} p_{j}^{(\gamma)}M^{(\gamma)} \right)
\label{Z1}
\ee
%
%
 
%%%%%%%%%%%%%%%%%%%%%%%%%%%%%%%%%%%%%%%%%%%%%%%%%%%%%%%%%%%%%%%%%%%%%%%
\begin{figure}[thb]
\begin{center}
\epsfig{file=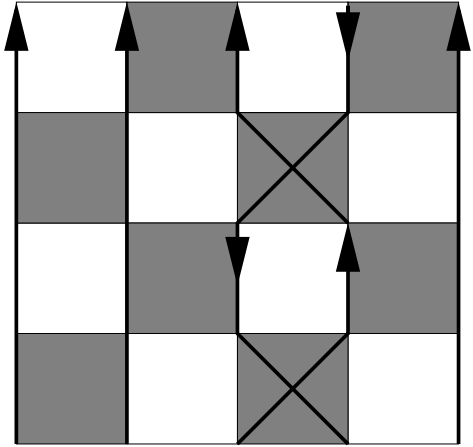,  height=6cm,
width = 6cm, angle=-90}
\medskip

\caption[]{Evolution of worldlines of up and down spins
in imaginary time. Periodic boundary conditions are assumed in 
both space and imaginary time. Note that we define worldlines
for both up- and down-spins, which differ by the direction
in imaginary time.}
\label{wordlines}
\end{center}
\end{figure}
%%%%%%%%%%%%%%%%%%%%%%%%%%%%%%%%%%%%%%%%%%%%%%%%%%%%%%%%%%%%%%%%%%%%

\begin{figure}
\begin{center}
\epsfig{file=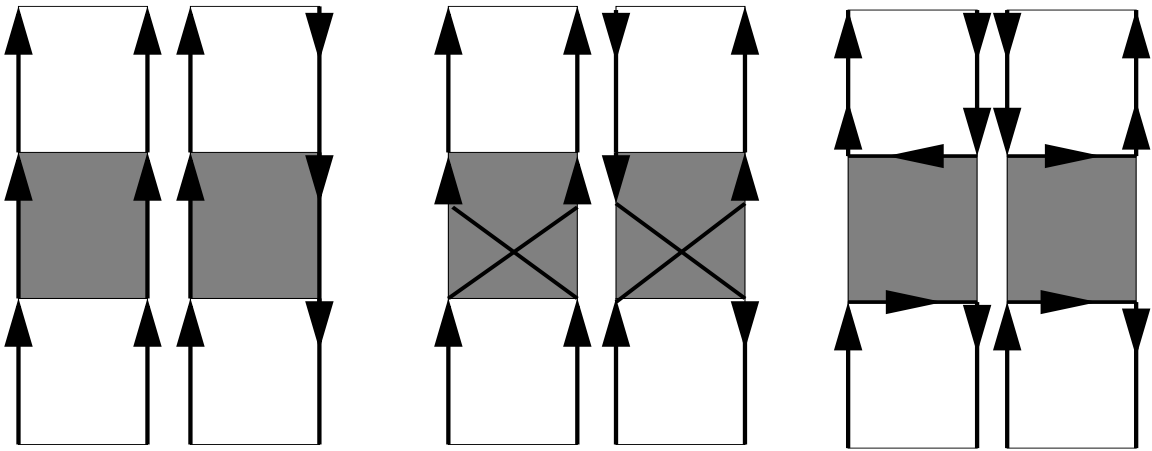,  height=15cm,
width = 6cm, angle=-90}
\medskip

\caption[]{Illustration of loop breakups, the directed lines 
represent the loop segments. From left to right the
vertical ($M^{(1)}$), diagonal ($M^{(2)}$) and the
horizontal breakup ($M^{(3)}$) are shown.}
\label{breakups} 
\end{center}
\end{figure}

Eq.\ (\ref{Z1}) can be interpreted in a geometrical way
(see Fig.\ \ref{wordlines}).
In the checkerboard picture the 
$M^{(\gamma)}$  matrices can be understood
as  different ways in which the worldlines 
can be broken in every plaquette and are usually called 
breakups.  
By taking one breakup per every plaquette we 
force the worldlines into closed paths which we call directed
loops (see Fig.\ \ref{breakups}).
A directed loop therefore follows the worldline of
an up-spin when it evolves in positive Trotter-time
direction and the world-line of a down-spin when it
evolves in negative Trotter-time direction.

In Fig.\ \ref{breakups} we show the graphic representation  
of the breakups $M^{(\gamma)}$. 
The lines now represent the directed loop segments.

%%%%%%%%%%%%%%%%%%%%%%%%%%%%%%%%%%%%%%%%%%%%%%%%%%%%%%%%%%%%%%%%%%%%%%%
%%%%%%%%%%%%%%%%%%%%%%%%%%%%%%%%%%%%%%%%%%%%%%%%%%%%%%%%%%%%%%%%%%%%

Eq.\ (\ref{Z1}) states that the partition function can
be obtained as a sum over all breakups. As a sum over all
breakups is equivalent to a sum over all loop configurations
$\{l\}$ we may rewrite Eq.\ (\ref{Z1}) as

\begin{equation}
Z=\sum_{\{l\}}\rho(\{l\})\
\Tr \prod_{n=1}^{N_T}
\bigotimes_{i=2m}M^{(\gamma_i)}
\bigotimes_{j=2m+1}M^{(\gamma_j)}~,
\label{Z2}
\end{equation}
where $\rho(\{l\})=\prod_i p_i^{(\gamma_i)}
                   \prod_j p_j^{(\gamma_j)}$.
Eq.\ (\ref{Z2}) leads to a very efficient 
MC-algorithm \cite{EVERTZ}: (a) Choose loop-breakups
$M^{(\gamma_i)}$ with probabilities
$p_i^{(\gamma_i)}$. (b) Construct the loop
configuration $\{l\}$ and flip all loops
with probability $1/2$. (c) Measure any
desired operator in all $2^{N_L(\{l\})}$
spin configurations reachable with 
independent loop flips (improved estimators),
where $N_L(\{l\})$ is the number of loops
in the loop configuration $\{l\}$.

%The matrix elements of the
%$M^{(\gamma)}$ are just one or zero determining
%which of the states on the plaquette is
%allowed and which not. They have the property
%
%%
%%
%\begin{equation}
%\langle\sigma_3,\sigma_4|M^{(1)}
%|\sigma_1,\sigma_2\rangle = \delta_{\sigma_1,\sigma_3}
%                            \delta_{\sigma_2,\sigma_4},\quad
%\langle\sigma_3,\sigma_4|M^{(2)}
%|\sigma_1,\sigma_2\rangle = \delta_{\sigma_1,\sigma_4}
%                            \delta_{\sigma_2,\sigma_3}~,
%\label{Vicente}
%\end{equation}
%%
%%
%%and 
%$\langle\sigma_3,\sigma_4|M^{(3)}
%|\sigma_1,\sigma_2\rangle = \delta_{\sigma_1,-\sigma_2}
%                            \delta_{\sigma_3,-\sigma_4}
%$.
%The trace over the product of transfer matrices
%$\Tr\,\prod_n\bigotimes_iM^{(\gamma_i)}
%         \bigotimes_jM^{(\gamma_j)}$
%occurring in Eq (\ref{Z2}) such determines which of
%the $2^{2LN_T}$ states on the Trotter lattice
%are allowed configuration for any given
%loop configuration $\{l\}$. As every loop can
%be flipped independently, there are 
%$2^{N_L(\{l\})}$ allowed spin configurations and 
%we may rewrite Eq.\ (\ref{Z2}) as
%
%%
%%
%\be
%Z= \sum _{\{l\}}\rho(\{l\})\ 2^{N_L(\{l\})}~.
%\label{Z3}
%\ee
%%
%
%Eq.\ (\ref{Z3}) represents the statistical model of 
%oriented loops which is equivalent to the original
%quantum mechanical model \cite{WIESE}. 

For later use we rewrite Eq.\ (\ref{Z2}) in a
form of traces over individual loops. 
Noting that vertical and diagonal loop segments
do not change the spin-direction
(see Fig.\ \ref{breakups}), we
may associate the $2\times2$ identity matrix
$\sigma^{0}=\left(\begin{array}{cc}1&0\\0&1\end{array}\right)$ 
with vertical and diagonal loop segments. As horizontal
loop segments do change the spin-direction, we
associate the Pauli-matrix
$\sigma^{x}=\left(\begin{array}{cc}0&1\\1&0\end{array}\right)$
with them. We then may rewrite
Eq.\ (\ref{Z2}) as

\be
Z= \sum _{\{l\}}\rho(\{l\})\,\prod_{l\in\{l\}}
\Tr_l \prod_{\mu} \sigma^{\gamma_\mu}~,
\label{Z4}
\ee 
where $\mu$ is an index running over loop $l$
and $\gamma_\mu=0,x$. $\Tr_l$ denotes the trace
over loop $l$.
Since $\Tr_l \prod_{\mu} \sigma^{\gamma_\mu}=2$,
Eq.\ (\ref{Z4}) is equivalent to a statistical mechanical
model of oriented loops,
$Z= \sum _{\{l\}}\rho(\{l\})\ 2^{N_L(\{l\})}$.

%%%%%%%%%%%%%%%%%%%%%%%%%%%%%%%%%%%%%%%%%%%%%%%%%%%%%%%%%
%%%%%%%%%%%%%%%%%%%%%%%%%%%%%%%%%%%%%%%%%%%%%%%%%%%%%%%%%%
  
\section{Correlation functions, improved estimators}
  
The expectation value of an operator ${\cal O}$ is

\be  
  \lan {\cal O} \ran  = \Tr ({\cal O} \exp(-\beta H) )=
\sum_{\alpha,\beta}  \lan \phi_{\alpha}|{\cal O}|\phi_{\beta}\ran
 \lan \phi_{\beta}|\exp(-\beta H)| \phi_{\alpha}\ran    
\label{Nondiagonal}     
\ee  

If ${\cal O}$ is diagonal in the basis $ \{|\phi_{\alpha} \ran \}$
then this procedure is straightforward.
The updating procedure generates a sequence of configurations 
$c_{i_{MC}}$ ($i_{MC}=1\dots N_{MC}$),
according with the distribution function of the system. 
In these configurations ${\cal O}$ takes a well defined 
value ${\cal O}(c_{i_{MC}})$, therefore:  

\be
\lan {\cal O}\ran =\frac{1}{N_{MC}}\sum_{i_{MC}}
{\cal O}(c_{i_{MC}})~. 
\label{MC_expect}
\ee

\begin{figure}
\begin{center}
\epsfig{file=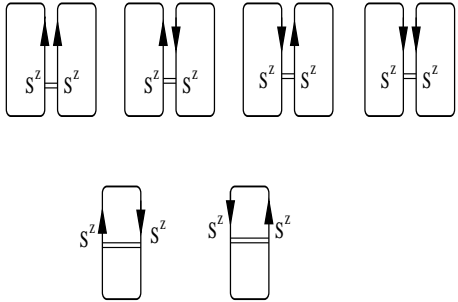,  height=6cm,
width = 4cm, angle=-90}
\medskip

\caption[]{Summing over all possible loop orientations, the
two-loop contributions cancel each other 
for $S^{z}_{\bf x} S^{z}_{\bf y}$. 
Only when the two spin-operators act 
on the same loop we get a non-vanishing contribution.}
\label{diagonal}
\end{center}
\end{figure}

The loop algorithm allows to measure an operator not only in 
$c_{i_{MC}}$
but in all configurations related by loop flippings. 
We illustrate the use of these improved estimators by computing 
${\cal O}=S^{z}_{\bf x} S^{z}_{\bf y}$ 
(here indices $\bf x$ and $\bf y$ label both
space and Trotter time (see Fig.\ \ref{diagonal}).
When $\bf x$ and $\bf y$ belong to different loops the orientation 
can be changed independently and the total contribution cancels.
By the contrary when $\bf x$ and $\bf y$ are  on the same loop 
the orientations of the loop in both sites are linked and 
these terms contribute for the two possible orientations of the loop.

%%%%%%%%%%%%%%%%%%%%%%%%%%%%%%%%%%%%%%%%%%%%%%%%%%%%%%%%%%%%%%%%%%%%%%%
%%%%%%%%%%%%%%%%%%%%%%%%%%%%%%%%%%%%%%%%%%%%%%%%%%%%%%%%%%%%%%%%%%%%

We will consider now  the problem of non diagonal operators. 
The expectation value of a  non diagonal operator 
${\cal O'}$ in the loop picture is, see Eq.\ (\ref{Z2}):   

\be    
 \lan {\cal O'} \ran=  \sum _{\{l\}}\rho(\{l\})\,
\Tr{\cal T}\left( {\cal O'} 
\prod_{n=1}^{N_T}
\bigotimes_{i}M^{(\gamma_i)}
\bigotimes_{j}M^{(\gamma_j)}
\right)~,
\label{Oprime} 
\ee  

where ${\cal T}()$ means proper imaginary time ordering.
Let us take as an example the two-point correlator 
${\cal O'}=S^{+}_{\bf x} S^{-}_{\bf y}$, 
    
%%%%%%%%%%%%%%%%%%%%%%%%%%%%%%%%%%%%%%%%%%%%%%%%%%%%%%%%%%%%%%%%%%%%%%%
\begin{figure}
\begin{center}
\epsfig{file=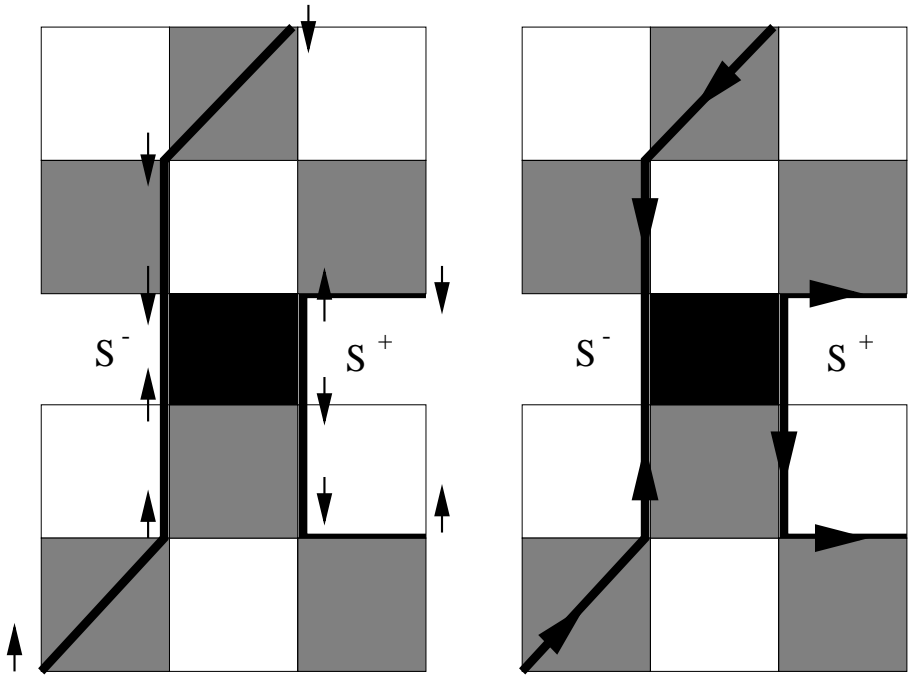,  height=12cm,
width = 9cm, angle=-90}
\medskip

\caption[]{The action of $S^{-}_{\bf x}S^{+}_{\bf y}$ 
can be represented as the insertion
of a new kind of plaquette, here depicted in black, which acts 
on two loop segments. This operator flips the spins and
changes therefore the orientation of the two loops
for the remaining segments. Left-picture: The arrows denote
the spin-direction. Right-picture: The arrows denote the
direction of the loops.}
\label{insert}
\end{center}
\end{figure}
%%%%%%%%%%%%%%%%%%%%%%%%%%%%%%%%%%%%%%%%%%%%%%%%%%%%%%%%%%%%%%%%%%%%
          
Graphically the evaluation  of an operator can be 
interpreted on the checkerboard framework
as the insertion of a new kind of plaquette. 
In Fig.\ \ref{insert} we show the action of that operator in the 
checkerboard picture. We note that an off-diagonal operator
in general reverses the direction of one or more loops.
The loop configurations generated by the MC
updating-procedure does, on the other hand, only generate
loops with well defined loop orientations.
Nevertheless there is a close connection between 
these two types of configurations which is
easy to understand in graphical terms.

%%%%%%%%%%%%%%%%%%%%%%%%%%%%%%%%%%%%%%%%%%%%%%%%%%%%%%%%%%%%%%%%%%
\begin{figure}
\begin{center}
\epsfig{file=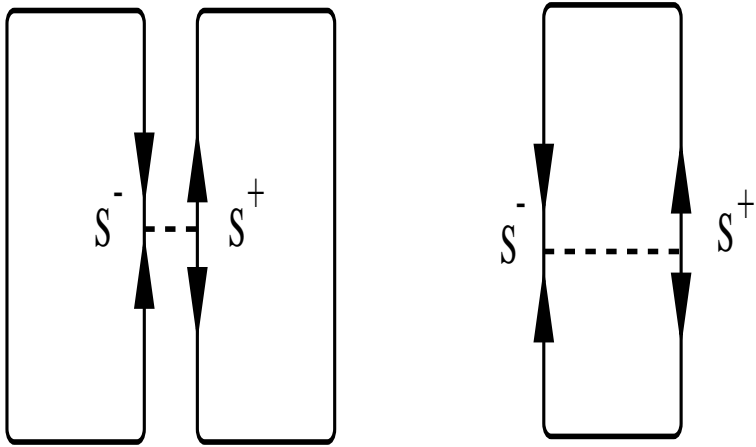,  height=10cm,
width = 6cm, angle=-90}
\medskip

\caption[]{
If $S^{+}_{\bf x}$ and $S^{-}_{\bf y}$ operate on different loops
none of them can be closed consistently in terms of loop 
orientation, represented here by arrows.
When both operators act on the same loop 
that configuration 
contributes to $\lan S^{+}_{\bf x} S^{-}_{\bf y} \ran$.}
\label{1loop}
\end{center}
\end{figure}
%%%%%%%%%%%%%%%%%%%%%%%%%%%%%%%%%%%%%%%%%%%%%%%%%%%%%%%%%%%%%%%%%%%%  

In Fig.\ \ref{insert} it is shown how the
flipping of one spin 'propagates' through  
the  loop, changing the orientation of the loop from that 
point. Thinking in terms of oriented loops it is 
obvious that with only one of these 
flipping processes  ($S_x^{+}$ or $S_y^{-}$) per loop,
it is not possible to
close the loop consistently. To reestablish the original loop orientation
it is necessary to have an  
even number of properly ordered  $S^{-}$ or $S^{+}$  
operators on the same loop  
to close it consistently in terms of loop orientation variables.
A loop which is not properly closed does not  contribute  
to $ \lan {\cal O'} \ran $. Then we can establish 
that for a two-point correlation function we only 
obtain a contribution when $x$ and $y$ belong to the 
same loop (see Fig.\ \ref{1loop}).
Eq.\ (\ref{Nondiagonal}) could suggest that measurements 
of non diagonal operators consume more 
computing time than diagonal operators, but
using this graphical picture we note that 
both computations can be implemented 
in an equivalent way.
     
These ideas can be justified in formal terms 
using Eq.\ (\ref{Z4}) and Eq.\ (\ref{Oprime}).
The $S^{+}_{\bf x}$ and $S^{-}_{\bf y}$ operators 
are placed in between of two $\sigma^{\gamma}$ matrices
belonging to neighboring plaquettes and
traces can be taken again independently in each loop.
We define the $2\times2$ matrices 
$\sigma^{+}=\left(\begin{array}{cc}0&1\\0&0\end{array}\right)$
and
$\sigma^{-}=\left(\begin{array}{cc}0&0\\1&0\end{array}\right)$.
For positive loop-direction (with respect to the Trotter
direction) $S^+$ is equivalent to $\sigma^+$, for
a directed loop segment with negative loop-direction
$S^+$ is equivalent to $\sigma^-$. For $S^-$ it is
just the other way round. The loop direction of relevance
here is the one before the insertion of either a
$S^+$ or a $S^-$ operator.

We start considering contributions to
$\langle S^{+}_{\bf x} S^{-}_{\bf y}\rangle$ where the
loop-direction at site $\bf x$ is up and down 
at site $\bf y$ (see Fig.\ \ref{1loop}).
The expectation value of the non-diagonal operator
$S^{+}_{\bf x} S^{-}_{\bf y}$ then becomes
(compare Eq.\ (\ref{Z4}))

\be
\langle S^{+}_{\bf x} S^{-}_{\bf y}\rangle
\rightarrow{1\over Z} \sum _{\{l\}}\rho(\{l\})\,
{\cal T}\left(\sigma_{\bf x}^{+} \sigma_{\bf y}^{+}
\prod_{l\in\{l\}}
\Tr_l \prod_{\mu} \sigma^{\gamma_\mu}\right)~.
\label{O4}
\ee 

Here ${\cal T}$ means proper time and space ordering.
When $\sigma^{+}_{\bf x}$ and $\sigma^{+}_{\bf y}$ 
are placed in different loops, the traces taken 
in these two loops cancel. If they are in the same loop 
the trace taken in that loop equals 1 (and not 2),
independently of the spin-configuration.
We will prove this last point now.
We start by writing the partial trace of the loop containing
$\sigma^{+}_{\bf x}$ and $\sigma^{-}_{\bf y}$ as

\[
T^{(++)} =\Tr_l\, \sigma_{\bf x}^{+} \left(\sigma^{x}\right)^{z_1} 
  \sigma_{\bf y}^{+} \left(\sigma^{x}\right)^{z_2}~,
\]
where we neglected the $\sigma^{0}$ matrices, as they are
just the identity matrices. We note that
$z_1+z_2$ is even since  
$\left(\sigma^{x}\right)^2=\sigma^{0}$
and because we are considering a loop which did
contribute to the partition function $Z$ before 
the $S_{\bf x}^+S_{\bf y}^-$ operators were inserted.
The $\sigma^x$ matrix corresponds to a horizontal
loop segment and such to a change in loop direction.
$z_1$ needs therefore to be odd 
(and therefore also $z_2$), since one needs an odd number
of directional inversions to arrive to a negative
loop direction at site $\bf y$, starting from a positive
direction at site $\bf x$. We may therefore rewrite
$T^{(++)}$ 
(using again  $\left(\sigma^{x}\right)^2=\sigma^{0}$) as

\[
T^{(++)} =\Tr_l\, \sigma_{\bf x}^{+} \sigma^x 
\sigma_{\bf y}^{+} \sigma^x \equiv1~,
\]
as one can easily evaluate. Similarly one can consider the
case when the initial loop directions are both positive
at sites $\bf x$ and $\bf y$. 
The expectation value of the non-diagonal operator
$S^{+}_{\bf x} S^{-}_{\bf y}$  becomes then in this case

\be
\langle S^{+}_{\bf x} S^{-}_{\bf y}\rangle
\rightarrow{1\over Z} \sum _{\{l\}}\rho(\{l\})\,
{\cal T}\left(\sigma_{\bf x}^{+} \sigma_{\bf y}^{-}
\prod_{l\in\{l\}}
\Tr_l \prod_{\mu} \sigma^{\gamma_\mu}\right)~.
\label{O5}
\ee 
The corresponding one-loop contributions then have the
form
\[
T^{(+-)} =\Tr_l\, \sigma_{\bf x}^{+} \left(\sigma^{x}\right)^{z_1} 
        \sigma_{\bf y}^{-} \left(\sigma^{x}\right)^{z_2}=
\Tr_l\, \sigma_{\bf x}^{+} \sigma_{\bf y}^{-}\equiv 1~,
\]
since both $z_1$ and $z_2$ have to be even in this case.
Similarly one can consider the two remaining cases
of loop directions down/up and down/down at the sites
$\bf x$ and $\bf y$. It is worthwhile noting, that one easily
proves along these lines the expected result
$\langle S^{+}_{\bf x} S^{+}_{\bf y}\rangle=0$.

%%%%%%%%%%%%%%%%%%%%%%%%%%%%%%%%%%%%%%%%%%%%%%%%%%%%%%
%%%%%%%%%%%%%%%%%%%%%%%%%%%%%%%%%%%%%%%%%%%%%%%%%%%%%% 

\section{General case n-point correlation functions}
    
In the last section we have shown how 
the loop orientation is the fundamental variable
to deal with the computation of 
correlation functions using improved estimators. 
In fact the problem of n-point correlation functions
can also be reduced to the study of
how the loop orientation is changed by the action 
of some operators.     
 
%%%%%%%%%%%%%%%%%%%%%%%%%%%%%%%%%%%%%%%%%%%%%%%%%%%%%%%%%%%%%%%%%%%%%%%
\begin{figure}[thb]
\begin{center}
\epsfig{file=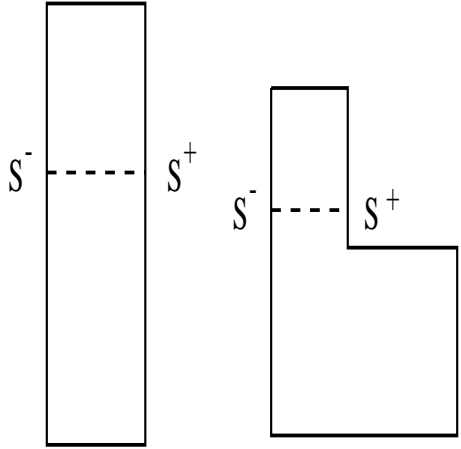,  height=6cm,
width = 6cm, angle=-90}
\medskip

\caption[]{Schematic example of two disconnected one-loop 
contributions to $\langle{\cal O}''\rangle$.
The dotted line in between two operators illustrates the
case of two operators at the same Trotter time.}
\label{2loopind}
\end{center}
\end{figure}
%%%%%%%%%%%%%%%%%%%%%%%%%%%%%%%%%%%%%%%%%%%%%%%%%%%%%%%%%%%%%%%%%%%%
             
\begin{figure}[thb]
\begin{center}
\epsfig{file=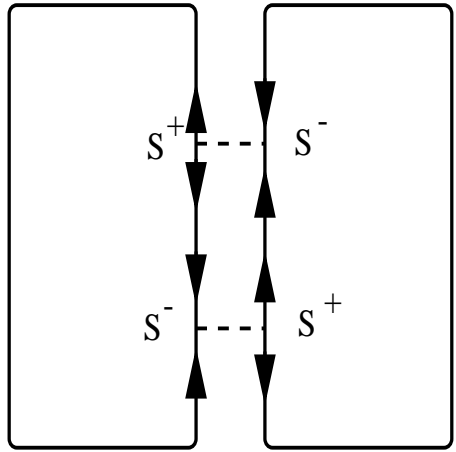,  height=6cm,
width = 6cm, angle=-90}
\medskip

\caption[]{Schematic example of connected two-loop contribution.
Two n.n.\ operators at the same Trotter time 
are connected with a dotted line.}
\label{2loopdep}
\end{center}
\end{figure}

We illustrate the case of two-loop terms for the
four-point correlation function  
${\cal O}''=S_{\bf x}^{+} S_{\bf y}^{-} 
S_{{\bf x}'}^{+} S_{{\bf y}'}^{-}$.
Here we consider the case relevant for the specific 
heat were  $({\bf x},{\bf y})$ and 
$({\bf x}',{\bf y}')$ are pairs of
real-space nearest neighbor (n.n.) sites at the same
Trotter time. 
This operator can generate several different kinds
of contributions. The first one is the case of
two disconnected one-loop contributions 
(see Fig.\ \ref{2loopind}).
This is the case if  $S_{\bf x}^{+}$ and $S_{\bf y}^{-}$
act in one loop and  
$S_{{\bf x}'}^{+}$ and $S_{{\bf y}'}^{-}$ in a second loop.
A second contribution arises if 
$S_{\bf x}^{+}$ and $S_{{\bf y}'}^{-}$
act in one loop and 
$S_{\bf y}^{-}$ and $S_{{\bf x}'}^{+}$ in a second loop
(see Fig.\ \ref{2loopdep}). We call this contribution
a connected two-loop contribution.
A third contribution arises when all four sites act on
the same loop. 

%%%%%%%%%%%%%%%%%%%%%%%%%%%%%%%%%%%%%%%%%%%%%%%%%%%%%%%%%%%%%%%%%%%%%%%
%%%%%%%%%%%%%%%%%%%%%%%%%%%%%%%%%%%%%%%%%%%%%%%%%%%%%%%%%%%%%%%%%%%

The evaluation of a single off-diagonal four-point
operator  ${\cal O}''$ does not pose a problem
within the loop algorithm. For the case of interest,
the specific heat a few additional points need to be
kept in mind. The specific heat $c_V$ is given by

\be
c_{V}= \frac{\beta^{2}}{L\,N_T^{2}}\left[
 \sum_{{\bf x},{\bf x}'}
\lan\, ( {\bf S}_{\bf x}\cdot{\bf S}_{\bf y})
       ( {\bf S}_{{\bf x}'}\cdot{\bf S}_{{\bf y}'})
\,\ran
-\left(\sum_{\bf x}
\lan {\bf S}_{\bf x}\cdot{\bf S}_{\bf y}\ran\right)^{2} 
\right]~,
\label{c_V}
\ee
where, again, $({\bf x},{\bf y})$ and
 $({\bf x}',{\bf y}')$ are pairs of (real-space) 
n.n.\ sites on the Trotter lattice.
The first term of Eq.\ (\ref{c_V})  
is a local energy-energy correlation function. When,
$\bf x$ and $\bf y$ belong to a loop and 
${\bf x}'$ and ${\bf y}'$ to another, we generate 
two-loop 
disconnected terms (as the one illustrated in
Fig.\ \ref{2loopind})
that can be computed from the 
expectation value of the internal energy, the 
second term of specific heat.
The energy in a given MC-configuration, $E_{i_{MC}}$,
can be written 
as a sum of the energy in the $N_{L}(i_{MC})$ loops in this
MC-configuration:  

\[ 
E_{i_{MC}}=\sum_{l=1}^{N_L(i_{MC})}E_{i_{MC}}^{l}~.
\]

\begin{figure}[thb]
\begin{center}
\epsfig{file=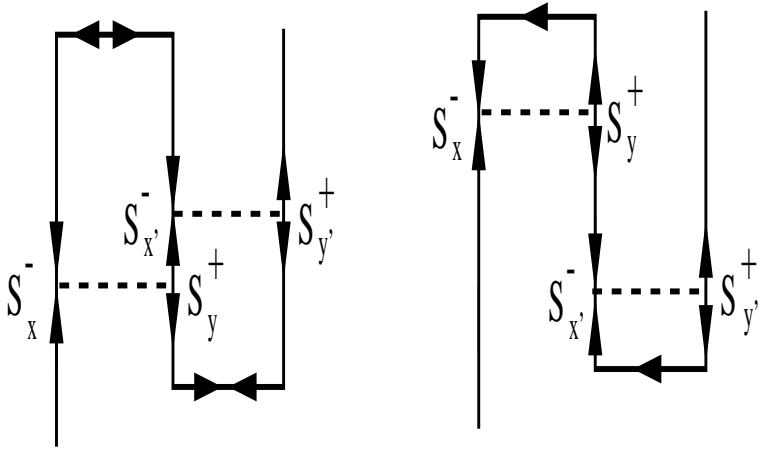,  height=10cm,
width = 6cm, angle=-90}
\medskip

\caption[]{On the left we show an example of 
non-contributing configuration to the
specific heat. The loop orientation is ill defined and 
therefore this configuration does not contribute. On the right 
we see a contributing one-loop configuration.  }
\label{no1loop}
\end{center}
\end{figure}

With this definition we obtain

\[ 
c_V^{(ind)}={1\over N_{MC}}\sum_{i_{MC}}
\sum_{l \neq k}E_{i_{MC}}^{l}E_{i_{MC}}^{k}
={1\over N_{MC}}\sum_{i_{MC}} \left[
\left(\sum_{l} E_{i_{MC}}^{l}\right)^{2}-
\sum_{l} \left(E_{i_{MC}}^{l}\right)^{2} 
\right]~,
\]
where $c_V=c_V^{(conn)}+c_V^{(ind)}$. 
For the evaluation of the connected term
$c_V^{(conn)}$ one has to evaluate the off-site terms,
$c_V^{(off)}$,
where the pairs $({\bf x},{\bf y})$ and 
$({\bf x}',{\bf y}')$ are disjunct,
separately from the on-site terms,
$c_V^{(on)}$, where they are not disjunct:
$c_V^{(conn)}=c_V^{(off)}+c_V^{(on)}$.
By spin-algebra the on-site terms reduce to
general two-point correlation functions.
The (connected) off-site contributions fall in
three categories, depending on the number $S^z$
operators involved (four, two or zero). The
contributions with four $S^z$ operators
have one and two loop contributions. A connected  term
with two $S^z$ operators has no two-loop contribution.
Every correlation with two $S^z$ operators has the form
$S^{z}_{\bf x}S^{z}_{\bf y}S^{+}_{{\bf x}'}S^{-}_{{\bf y}'}$.
If the indices ${\bf x}$ and ${\bf y}$
are not in the  same loop the two $S^{z}$ operators act 
in different loops and their traces  
cancel for the reason explained in section III.
The same reasoning is valid for ${\bf x}'$ and ${\bf y}'$   
with the operators $S^{+}$ and $S^{-}$.
Finally, terms with no $S^{z}$ operators can have 
two loop contributions (see Fig.\ \ref{2loopdep})
and also one-loop contributions when the 
$S^{+}$ and $S^{-}$ are properly ordered 
along the loop to close the loop coherently 
in terms of loop orientation. On the left of Fig.\ \ref{no1loop}
we see that an arbitrary insertion of the operators
$S^{+}$ and $S^{-}$ can produce a conflict 
on the orientation of the loop. Technically, 
the value of the trace taken along the loop will depend   
on the  structure of the correlator. This structure determines the order
of the insertion of the  $\sigma^{+}$ and  $\sigma^{-}$ matrices.
For example the trace along the loop on the left of  
Fig.\ \ref{no1loop} is: 
\[ 
\Tr(\sigma^{-}_{\bf x}  \sigma^{x}  \sigma^{-}_{{\bf x}'}
\sigma^{-}_{\bf y}\sigma^{x}\sigma^{+}_{{\bf y}'})=0~.
\]
For the loop on the right of Fig.\ \ref{no1loop} it is:
\[ 
\Tr(\sigma^{-}_{\bf x}\sigma^{x}\sigma^{-}_{\bf y}
\sigma^{+}_{{\bf x}'}\sigma^{x}\sigma^{+}_{{\bf y}'})=1~.
\]
%
%  
    
%%%%%%%%%%%%%%%%%%%%%%%%%%%%%%%%%%%%%%%%%%%%%%%%%%%%%%%%%%%%%%%%%%%%%%%
%%%%%%%%%%%%%%%%%%%%%%%%%%%%%%%%%%%%%%%%%%%%%%%%%%%%%%%%%%%%%%%%%%%%

\begin{figure}[thb]
\begin{center}
\epsfig{file=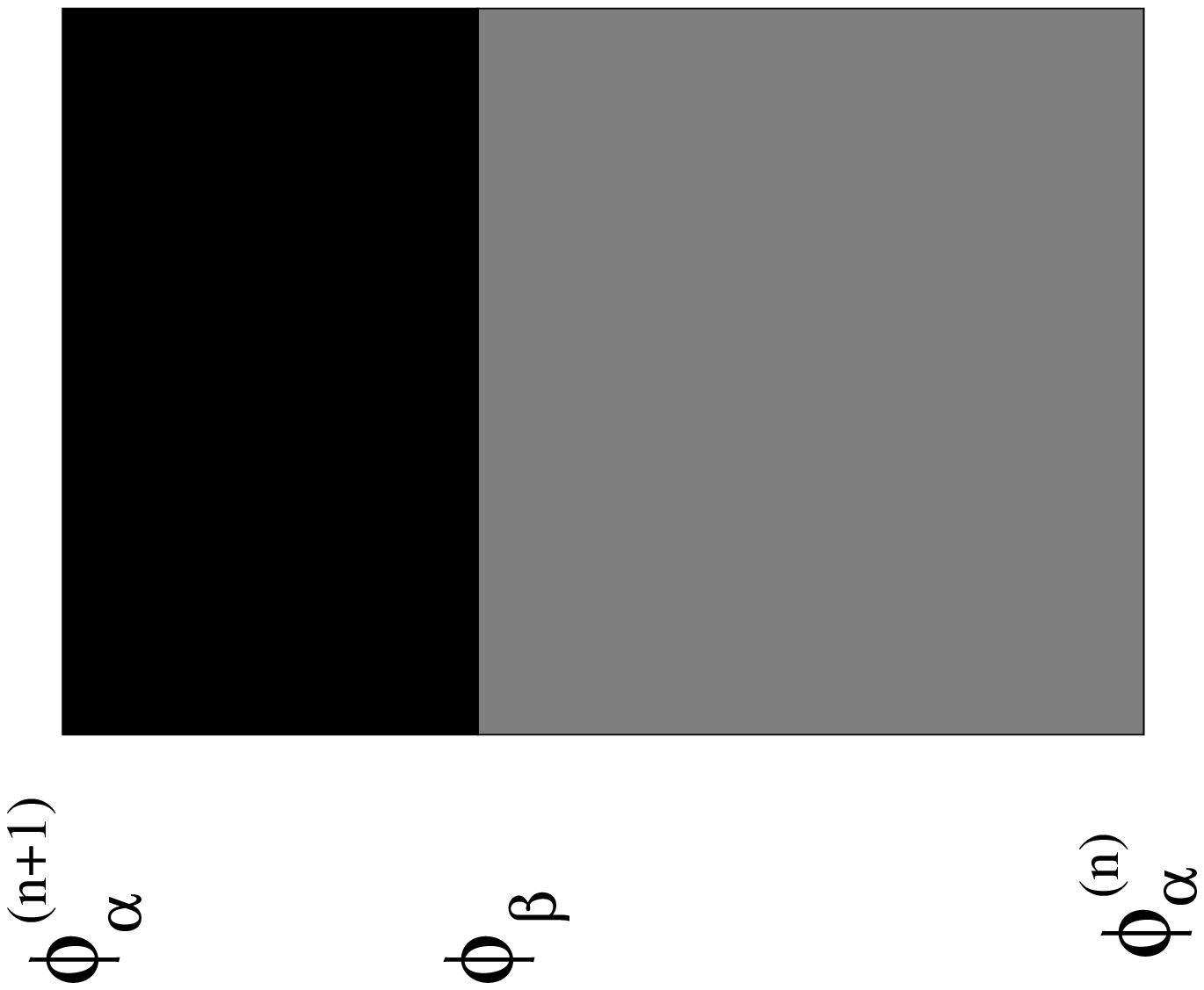,  height=6cm,
width = 6cm, angle=-90}
\medskip

\caption[]{New structure of the plaquette in the reweighting method.
The grey plaquette is the conventional plaquette where the 
evolution in imaginary time takes place, the black plaquette represents 
the operator $\cal{O}$ on the basis of $\sigma_{z}$. The product 
of the two matrices generates a new composite plaquette where 
the new weight is defined.}
\label{reweight}
\end{center}
\end{figure}

It is possible to evaluate certain off-diagonal operators $\cal O$ 
by an alternative method. The condition is, that the
operator can be expressed by a sum of local operators which
do involve the same pairs of sites $\lan l,l'\ran$ as the
Hamilton-operator $H=\sum_{\lan l,l'\ran} H_{l,l'}$:
${\cal O}=\sum_{\lan l,l'\ran} {\cal O}_{l,l'}$.
It is then possible to compute  $\lan \cal O\ran$ by a reweighting
method. The idea is to extend the plaquette of the 
checkerboard representation by new internal degrees of
freedom, $\sum_{\beta} |\phi_{\beta}\ran \lan \phi_{\beta}|$
(see Fig.\ \ref{reweight}). The reweighted matrix element
of $\lan {\cal O}_{x,x'} \ran$ is then
\be
{\cal O}_{\alpha_n,\alpha_{n+1}}^{(n)}(x,x') =  
\sum_{\beta}\,{
\lan  \phi_{\alpha_n}^{(n)}|{\cal O}_{x,x'}|\phi_{\beta}\ran
\, \lan \phi_{\beta}|\exp(-\Delta\tau H_{x,x'})
 |\phi_{\alpha_{n+1}}^{(n+1)} \ran\over
 \lan \phi_{\alpha_n}^{(n)}|\exp(-\Delta\tau H_{x,x'})
 |\phi_{\alpha_{n+1}}^{(n+1)}\ran
	      } ~,
\ee  
where $x$ and $x'$ denote combined space-time indices.
For a given spin-configuration 
$c_{i_{MC}}=\{\phi_{\alpha_n}^{(n)}|(n=1,\dots,N_T)\}$ 
the
off-diagonal expectation value of ${\cal O}(c_{i_{MC}})$ is
${\cal O}(c_{i_{MC}})=1/N_T\sum_{\lan x,x'\ran,(n)}
{\cal O}_{\alpha_n,\alpha_n+1}^{(n)}(x,x')
$ and 
$\lan {\cal O}\ran=1/N_{MC}\sum_{i_{MC}} {\cal O}(c_{i_{MC}})$
(see Eq.\ (\ref{MC_expect})).

%%%%%%%%%%%%%%%%%%%%%%%%%%%%%%%%%%%%%%%%%%%%%%%%%%%%%%%%%%%%%%%%%%%%%%%
%%%%%%%%%%%%%%%%%%%%%%%%%%%%%%%%%%%%%%%%%%%%%%%%%%%%%%%%%%%%%%%%%%%%

The reweighting method  may also be applied to
specific heat, which is the sum of products of local operators. 

From the point of view of the complexity of the algorithm,
measuring four-point correlation functions
requires more computing time than 
two-point correlation functions. 
For the  latter is only necessary  to know whether or not 
two sites are in the same loop. This information 
can be obtained at the same time the loop is constructed 
and consequently the computing time remains proportional 
to $LN_T$. For n-point correlation functions the situation
is more complex. In this case, there are contributions
involving two or more loops and at the same time 
non-diagonal operators give different contributions 
depending on  how they are ordered on the loop.
In practice this depends on the shape of the loops 
in each configuration.   
A rigorous study of the performance of the method
must include an analysis of the behavior of the 
statistical errors as a function of the
temperature, size, number and type 
of operator involved in the correlation functions
and the details of the Hamiltonian. 
This detailed analysis of technical aspects 
of n-point correlations will be presented elsewhere.
      
\section{Results}

As an application of the rules explained in this paper
we have computed the specific
heat for a Heisenberg chain and for a ladder with
$J_{\perp}=0.5J$ (which corresponds to the ratio
for the ladder-compound Sr$_{14}$Cu$_{24}$O$_{41}$
\cite{Dagotto_Rice}).

%%%%%%%%%%%%%%%%%%%%%%%%%%%%%%%%%%%%%%%%%%%%%%%%%%%%%%%%%%%%%%%%%
\begin{figure}[thb]
\begin{center}
\epsfig{file=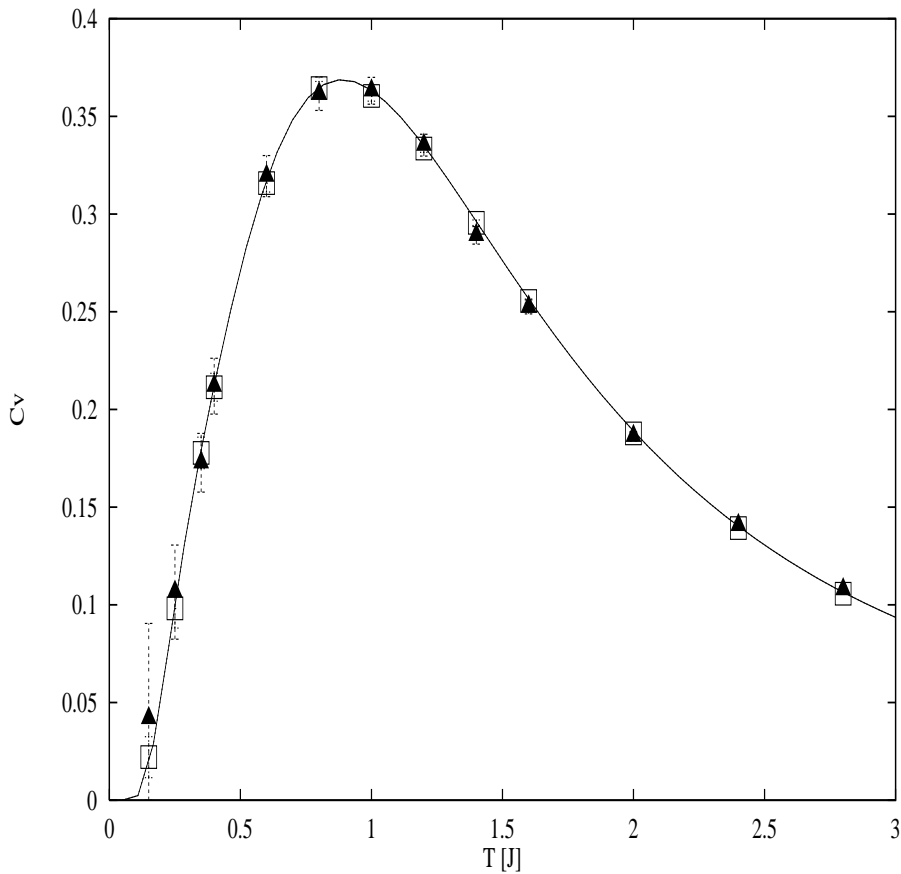,  height=12cm,
width = 12cm, angle=-90}
\medskip

\caption[]{Specific heat for the Heisenberg model in the 8-site chain
with $J=2$, as a function of temperature in units of $J$.
The two sets of data correspond to the QMC simulations with 
improved estimators (open squares) and with the reweighting 
method (filled triangles), for the same number of QMC steps.  
The solid line is the exact diagonalization data.
}
\label{L_8}
\end{center}
\end{figure}
%%%%%%%%%%%%%%%%%%%%%%%%%%%%%%%%%%%%%%%%%%%%%%%%%%%%%%%%%%%%%%%%%%

In  Fig.\ \ref{L_8} we compare exact diagonalization results
with the results using  the method described above and the reweighting method 
for the same number of
MC steps. The error bars in these two methods 
are also compared. For the lowest temperature 
the error bar with improved estimators are 
6 times smaller. Taking into account that 
error bars decay as $\frac{1}{\sqrt{N_{MC}}}$
we expect that without using improved estimators 
36 times  more MC steps are necessary to get 
equal size error bars. The statistical errors  
are amplified by the factor $\beta^2$. This factor
and the substraction of similarly large numbers
lead to large error bars at low temperatures. 

\begin{figure}[thb]
\begin{center}
\epsfig{file=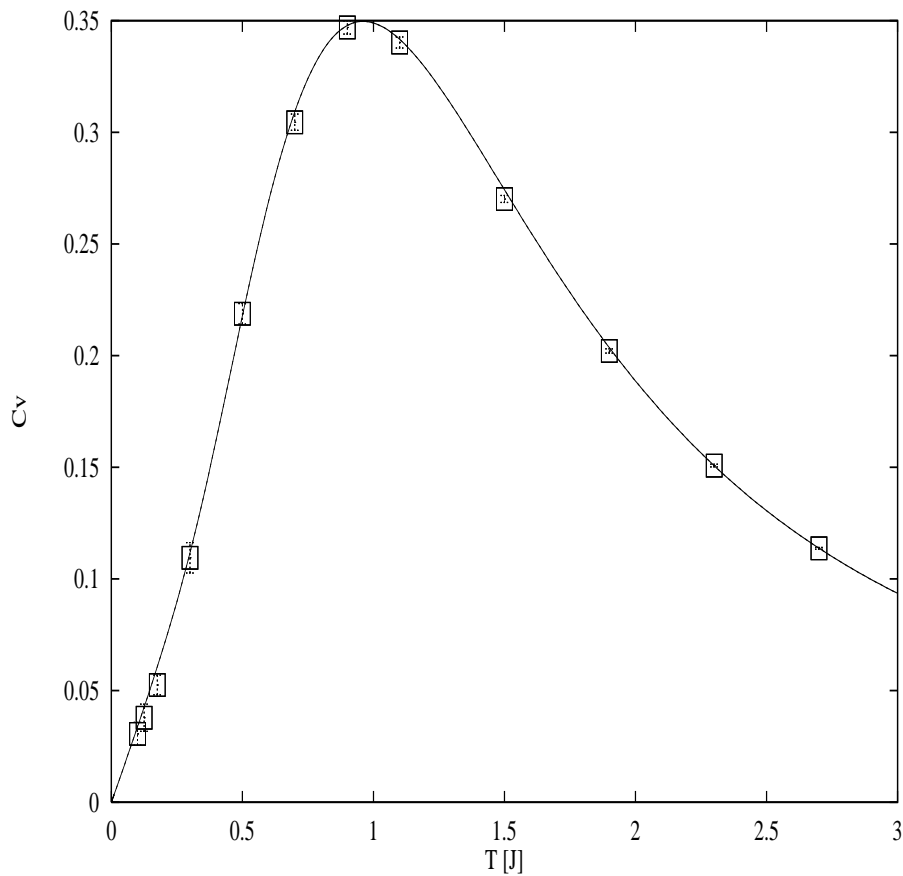,  height=12cm,
width = 12cm, angle=-90}
\medskip

\caption[]{ Specific heat for a 100 sites Heisenberg with J=2.0
chain using improved estimators, as a function of temperature,
in units of $J$. The error-bars are smaller than the symbol
sizes. The solid line is the
exact Bethe-Ansatz result for the infinite-chain 
\protect\cite{Kluemper}.}
\label{L_100}
\end{center}
\end{figure}

In the Fig.\ \ref{L_100} we present results
for the  specific heat of a 100-site Heisenberg
chain. To reproduce the linear regime at low 
temperatures it is necessary to perform a careful
extrapolation to $\Delta \tau \go 0 $ taking 
half a million of MC steps for each  $\Delta \tau$
values and 10 different  $N_T$ values  
ranging from 20 to 200.

%%%%%%%%%%%%%%%%%%%%%%%%%%%%%%%%%%%%%%%%%%%%%%%%%%%%%%%%%%%%%%%%%%
%%%%%%%%%%%%%%%%%%%%%%%%%%%%%%%%%%%%%%%%%%%%%%%%%%%%%%%%%%%%%%%%%% 

\begin{figure}[thb]
\begin{center}
\epsfig{file=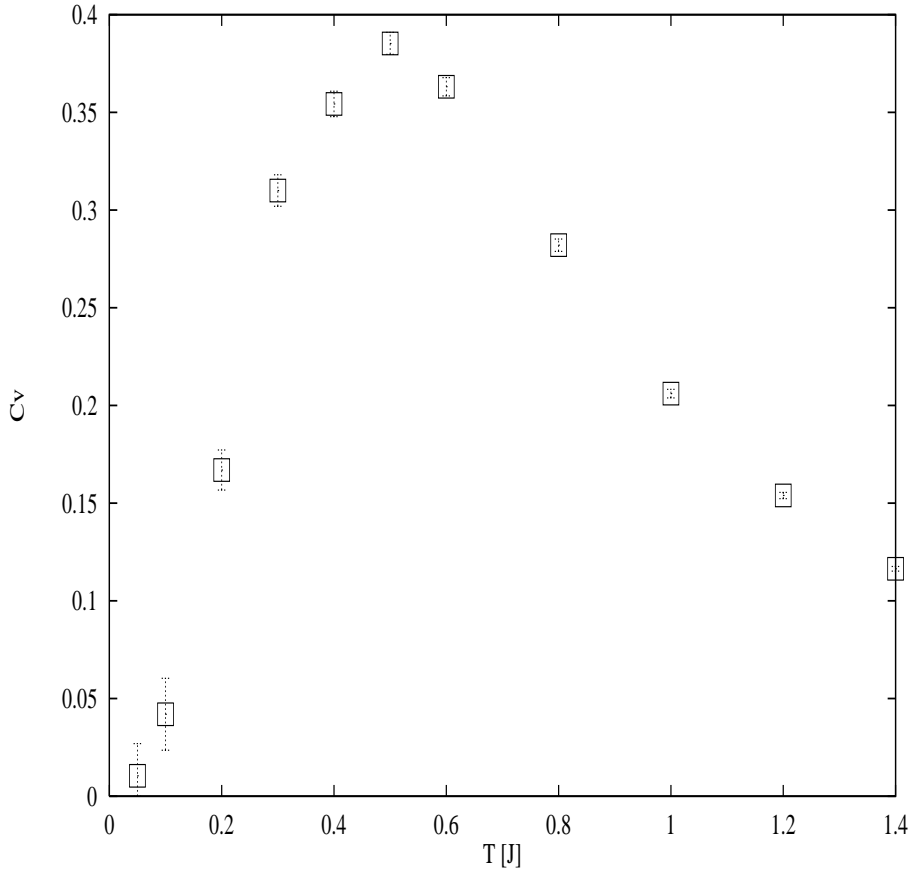,  height=12cm,
width = 12cm, angle=-90}
\medskip

\caption[]{Specific heat for the $2\times201$ ladder with 
twisted boundary conditions as a function of temperature
in units of $J$.
The values of the couplings are  $J=1.0$  
and $J_{\perp}=0.5J$.}
\label{L_2x201}
\end{center}
\end{figure}

In Fig.\ \ref{L_2x201} we present results for the
 two-leg ladder of $2\times201$ sites with twisted
boundary conditions (i.e.\ for $J_{\perp}=0$ this
system corresponds to a L=402-site Heisenberg chain).

%%%%%%%%%%%%%%%%%%%%%%%%%%%%%%%%%%%%%%%%%%%%%%%%%%%%%%%%%%%%%%%%%%
%%%%%%%%%%%%%%%%%%%%%%%%%%%%%%%%%%%%%%%%%%%%%%%%%%%%%%%%%%%%%%%%%%%%  

\section{Conclusions}

We have presented detailed rules on how to evaluate general,
off-diagonal n-point Greens functions within the
loop algorithm. These rules have a very simple interpretation
in the picture of oriented loops. They state that the
loop-orientation has to close coherently whenever a certain
number of non-diagonal operators are inserted. We have shown
how to apply these rules to the case of the specific heat and
presented results for the 1D-Heisenberg model and a
ladder system.

\section{Acknowledgments}

We would like to acknowledge discussions with
Matthias Troyer,  Naoki Kawashima and Andreas
Kl\"umper and the support of the German Science
Foundation. We acknowledge the hospitality of the
ITP in Santa Barbara. This research was supported
by the National Science Foundation under Grant
No. PHY94-07194.

\vfill\eject

\end{document}